# Systematic Synthesis and Design of Ultra-Low Threshold Parametric Frequency Dividers

Hussein M. E. Hussein, *Student Member, IEEE,* Mahmoud A. A. Ibrahim, *Student Member, IEEE,* Giuseppe Michetti, *Student Member, IEEE,* Matteo Rinaldi, *Member, IEEE,* Marvin Onabajo, *Senior Member, IEEE,* and Cristian Cassella, *Member, IEEE*

*Abstract*—A new method is discussed for the systematic synthesis, design and performance optimization of varactor-based parametric frequency dividers (PFDs) exhibiting an ultra-low power threshold ($P_{th}$). For the first time, it is analytically shown that the $P_{th}$-value exhibited by any PFD can always be expressed as an explicit closed-form function of the different impedances forming its network. Such a unique and unexplored property permits to rely on linear models, during the PFD design and performance optimization. The validity of our analytical model has been verified, in a commercial circuit simulator, through time-domain and frequency-domain algorithms. To demonstrate the effectiveness of our new synthesis approach, we also report on a lumped prototype of a 200:100MHz PFD, realized on a printed circuit board (PCB). Although inductors with quality factors lower than 50 were used, the PFD prototype exhibits a $P_{th}$-value lower than −15dBm. Such a low $P_{th}$-value is the lowest one ever reported for passive varactor-based PFDs, operating in the same frequency range.

*Index Terms*—Frequency Dividers, Linear-Time-Variant (LTV) System, Auxiliary Generators, Nonlinear Dynamics

## I. Introduction

IN the last decades, a growing attention has been paid to the development of new electronic components leveraging strong nonlinear dynamics in order to surpass the limitations of the currently available devices and systems [1]–[12]. In particular, many research groups have looked at the possibility of exploiting nonlinear phenomena to attain frequency synthesizers (FSs) with record-low jitter levels [9], [13]–[20]. Only recently, one of the investigated approaches produced a new CMOS-compatible component referred to as parametric filter (PFIL) [21], [22], which shows the unprecedented ability to act as a non-autonomous jitter-cleaner, not requiring the use of a voltage-controlled-oscillator (VCO). PFILs leverage the complex nonlinear dynamics exhibited by varactor-based parametric frequency dividers (PFDs) [23]–[26], placed in non-autonomous feedback loops and directly connected at the output of a noisy FS. Despite the fact that a PFIL prototype showing a large phase-noise suppression was recently demonstrated [22], such system is characterized by a power consumption that is not suitable for low-power integrated electronics. For this reason, new strategies to reduce the power consumed by PFILs are required to enable their adoption in low-power systems. This critical power limitation is mostly determined by the minimum input power ($P_{th}$) that makes PFDs able to operate in their division regime. PFDs are nonlinear circuits that rely on the adoption of largely modulated reactances to activate a frequency division mechanism.

Because of their strong nonlinear behavior, the design of PFDs, through commercial circuit simulators, presents several challenges that have prevented achieving PFDs exhibiting ultra-low $P_{th}$-values [16], [24]. For instance, as these devices can exhibit abrupt changes in their electrical characteristics, the use of time-domain (TD) algorithms to model their response is only limited to PFDs using a reduced number of components. This constraint impedes to attain optimized PFD designs, with minimum $P_{th}$-values, through TD-based methods. In contrast, the detection of sub-harmonic oscillations through conventional Harmonic-Balance (HB) algorithms shows severe limitations due to the absence of sub-harmonic frequencies among those used to find the steady-state solution of time-varying circuits. Consequently, alternative approaches were developed to detect the onset of sub-harmonic oscillations, in PFDs, through perturbation methods [12], [27] or through the iterative determination of the conversion matrix [28], [29] associated to any adopted variable reactance. Unfortunately, all the previously developed approaches require the adoption of a fine sweep of an input parameter, such as the excitation frequency ($f_{pump}$), the input voltage ($V_1$ shown in Fig. 1) or the input power ($P_{in}$), to render HB-simulators able to reliably identify non-trivial dividing solutions. Hence, their use generally comes with a high simulation complexity that often makes them unsuitable when the goal is to design PFDs with minimized $P_{th}$-values.

Recently, while investigating the optimum design conditions to build nonreciprocal RF filters [30], through a network of modulated reactances, we discovered that it is always possible to express the transfer function describing the operation of such systems as an explicit function of the static equivalent impedance seen by each modulated component. The discovery of such a unique property led to augmented synthesis capabilities that allowed us to unveil the main design criteria and functionalities of a novel nonreciprocal RF component, with optimized architecture, insertion-loss and isolation. Here, we show that a similar property exists for PFDs, relating the stability of large-signal periodic regimes to the different static impedances forming their network. For the first time, a closed-form expression of the $P_{th}$-value exhibited by any 2:1 varactor-based PFD is derived and reported. In particular, we show that this key performance parameter can indeed be expressed as an explicit function of the impedances seen by the variable reactance towards the PFD circuit ports (input and output), and relative to the input (or pump, $f_{pump}$) and output ($f_{out}$) frequencies. Therefore, we demonstrate that the mini-mization of $P_{th}$ can be tackled through standard impedance synthesis approaches, thereby not requiring the adoption of perturbation-based or iterative techniques that are often hard to be used, without increasing the design and the simulation complexity. This complexity can even be unsustainable when targeting ultra- and super-high-frequency (UHF/SHF) PFDs, whose design requires electromagnetic simulations to account and compensate for the significant parasitics generally introduced by the board layout. Thanks to the derived closed-form expression of $P_{th}$, a new design guideline for ultra-low threshold PFDs is unveiled and reported, hence providing the means to finally achieve low-power PFILs. To demonstrate the validity and effectiveness of our findings, a 200:100MHz PFD using lumped off-the-shelf components was designed and built on a printed circuit board (PCB). Even though this device uses inductors with quality factors ($Q$) lower than 50, the engineered strategic selection of its passive components renders it able to achieve a $P_{th}$-value lower than -15dBm. To the best of the authors knowledge, this value is the lowest one ever reported for passive PFDs, operating in the same frequency range. [16], [25], [28], [31].

## II. DETECTION OF SUB-HARMONIC OSCILLATIONS IN PARAMETRIC FREQUENCY DIVIDERS

The detection of parametric instabilities represents a significant challenge for most commercial circuit simulators [24]. In particular, a reliable identification of sub-harmonic oscillations would not only enable optimal performance in parametric circuits, but would also allow to prevent drops of spectral purity and power-efficiency in other circuit components, such as amplifiers and frequency multipliers [32], [33]. Several research groups have looked at possible approaches to identify the rising of parametric oscillations in RF systems. Some approaches use TD-algorithms. However, due to the abrupt functional changes occurring at points of marginal stability, the use of TD-based detection methods to analyze the operation of parametric circuits may lead to severe convergence issues. These problems can be only overcome through the adoption of finer time steps, which frequently implies unsustainable computation times. For this reason, their use can even be impossible when analyzing complex systems, such as PFILs or more advanced PFD designs. On the other hand, when using frequency-domain (FD) based algorithms, most commercial HB-circuit simulations do not detect the onset of oscillations occurring at frequencies that are sub-multiples of any input frequency, since sub-harmonic frequencies are not included among those used by these simulators to evaluate currents and voltages, in analyzed circuits. However, as FD-methods permit to efficiently characterize the behavior of any circuit, through a much shorter computation time than TD-methods, enabling their use is key to most efficiently design PFDs. So, different approaches have been explored to achieve a reliable behavioral prediction of parametric components and systems through HB-methods. In particular, in [34], a voltage-auxiliary-generator (AG) technique was developed to extract the $P_{th}$-value attained by varactor-based PFDs. This approach is based on the artificial introduction of a voltage generator in series with an ideal frequency selective resistive filter, and placed in parallel to the adopted modulated reactance. This generator, which is characterized by an excitation frequency equal to the divided output frequency ($f_{out}$), applies a low-voltage signal in the circuit, thus forcing any HB-simulator to consider a signal at $f_{out}$ during its computation. The signal generated by the AG acts as noise, characterized by an impulsive frequency distribution centered at $f_{out}$. Its use permits to assess the PFD stability, at $f_{out}$, as the amplitude of the main excitation voltage, at $f_{pump}$ (i.e., $2f_{out}$), is increased. Although this method permits to detect the rising of parametric instabilities, it requires iterative and time-consuming simulations to find the steady-state response of PFDs, after these systems transition into their division region (i.e., after the onset of the parametric instability). Consequently, the use of AGs also comes with an excessively long computational time when intricate optimizations are needed to find the best PFD design. As an alternative approach, a novel detection technique has recently been developed [16]. This is based on the introduction of a power auxiliary generator (pAG), in the PFD output mesh and on behalf of the PFD output load ($R_L$). A pAG is characterized by an ideal voltage generator, at $f_{out}$, in series with its internal impedance ($Z_g$, set to be equal to $R_L$). The available power ($P_{sub}$) of the pAG is kept small enough to ensure that no perturbation of the circuit behavior is generated as a result of its use. Thus, the introduction of the pAG allows to include $f_{out}$ in the list of frequencies used by any HB simulator, without perturbing the impedances seen by any modulated reactance in the circuit. Furthermore, contrary to the AG, the use of a pAG avoids to rely on time-consuming optimizations to extract the steady-state response of PFDs. In fact, when a pAG is used, the PFD output voltage can be directly extracted from the HB-simulated voltage across the pAG, at $f_{out}$. Such voltage, automatically differs from the originally set value, corresponding to $P_{sub}$, after the start of sub-harmonic oscillations in the circuit. This unique feature enables the direct extraction of the PFD output spectrum even for $P_{in} > P_{th}$. However, in order to reliably use the discussed pAG-technique, very fine sweeps of specific controlling parameters, such as $P_{in}$ or $f_{pump}$, must still be implemented to facilitate the HB-convergence to non-trivial dividing solutions. This feature also makes the pAG-technique non-ideal when optimizing PFD designs targeting ultra-low $P_{th}$-values. Hence, when the minimization of $P_{th}$ is the main design objective, gaining intuition about the different factors affecting its value is fundamental to overcome the limits of previously developed approaches. To do so, one of objectives of this work has been to compute a generic closed-form expression that can be easily accessed through linear simulation algorithms to estimate and minimize the $P_{th}$-value exhibited by any PFD that is to be designed.

## III. AN ANALYTICAL INVESTIGATION OF $P_{th}$ IN PFDS

As we are interested in identifying the critical parameters influencing the dynamics of PFDs, we can use analytical methods to determine $P_{th}$. For simplicity, we start our analysis by using a simplified PFD circuit (Fig. 1) based on a T-network topology. This circuit includes two static series resonant LC-tanks ($L_1 - C_1$ and $L_2 - C_2$, in series to the input or output terminals, respectively) and a shunt modulated LC-tank relying on an inductor ($L_3$) with a biased varactor ($C_3(v(t), t)$, where $v(t)$ and $t$ are the voltage across the varactor and the time respectively). Next, we extract the TD system of the second-order differential equations (1) mapping the charge $q_1(t), q_2(t)$ and $q_3(t)$) entering the only circuit node ($N1$) of the circuit. Analyzing these equations permits to define an equivalent mono-lateral FD system for more general PFD-designs relying on generic impedances ($Z_1$, $Z_2$ and $Z_3$) instead of the initially considered LC-tanks (Fig. 1). As discussed in the following section, the analysis of this transformed problem in a FD representation permits to extract $P_{th}$ by looking at the stability of the trivial solution vs. the applied input power ($P_{in}$). To verify the validity of our findings, the extracted $P_{th}$-value, for a specific PFD design (used as a case study), is then compared with the corresponding values found through numerical TD- and FD-methods. After identifying the conditions to start sub-harmonic oscillations in the circuit, we also describe an analytical approach to estimate the steady-state response of the same investigated PFD, at half of the driving frequency and for $P_{in}$-values exceeding $P_{th}$.

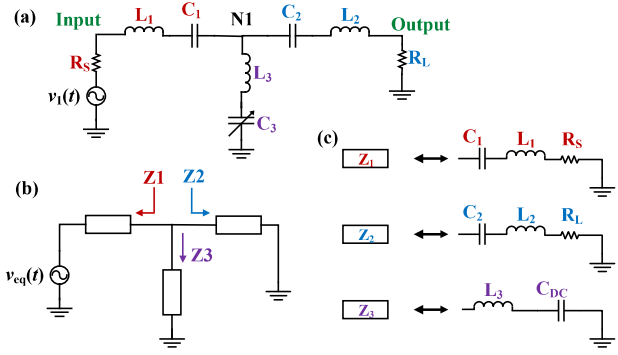

Fig. 1. (a) Initially analyzed PFD design, which is formed by three LC-tanks (2 static linear ones and one including a modulated varactor) and two resistors mapping the source resistance and load resistance, respectively. This PFD circuit is initially studied as the means to extract an equivalent frequency-domain (FD) model that can be used to investigate the behavior of more general PFD designs. (b) A more general PFD circuit model that includes three general impedances ($Z_1$, $Z_2$ and $Z_3$) on behalf of the LC-tanks used in a, which is adopted in our FD analysis reported in this work. (c) Schematic of $Z_1$, $Z_2$ and $Z_3$ for the circuit in a.

### A. A Closed-Form Expression for $P_{th}$

We start our analysis from the simplified circuit shown in Fig. 1-a. By using state variables for the charge $q_1(t), q_2(t)$ and $q_3(t)$ in the capacitors $C_1, C_2$ and $C_3$, respectively, we can write the system of Kirchhoff's equations that describe their evolution over time (1), when assuming zero pre-stored electrical energy in all lumped components of the circuit. $R_S$ and $R_L$ represent the source and load impedances, respectively, and $v_1(t)$ is a continuous wave input signal characterized by a peak level equal to $V_1$.

$$v_1(t) = \frac{q_1(t)}{C_1} + \frac{q_2(t)}{C_2} + R_S q_1'(t) + R_L q_2'(t) + L_1 q_1''(t) + L_2 q_2''(t)$$
$$v_1(t) = \frac{q_1(t)}{C_1} + \frac{q_3(t)}{C_3} + R_S q_1'(t) + L_1 q_1''(t) + L_3 q_3''(t)$$
$$q_1(t) - q_2(t) - q_3(t) = 0 \quad (1)$$

It is important to point out that $C_3$ in (1) is a voltage-dependent nonlinear capacitor. Thus, in order to emphasize its time-varying characteristic, $C_3$ will be indicated as $C_3(t)$.

$$C_3(t) = C_{DC}\left(1 + \epsilon \frac{C_d q_3(t)}{C_{DC}} + \epsilon^2 \frac{C_{d2} q_3(t)^2}{C_{DC}^2}\right) \quad (2)$$

For simplicity, $C_3(t)$ can be approximated with its second-order Taylor expansion (see (2)) around its average DC component ($C_{DC}$). In (2), $\epsilon$ represents an arbitrary small real parameter that is used to control the perturbation order [35], [36] adopted for $C_3(t)$, at different stages of our analytical treatment. After substituting (2) into (1) and in the limit of small perturbations ($\epsilon \to 0$), equation (1) can be rewritten as:

$$v_1(t) = \frac{q_1(t)}{C_1} + \frac{q_2(t)}{C_2} + R_S q_1'(t) + R_L q_2'(t) + L_1 q_1''(t) + L_2 q_2''(t)$$
$$v_1(t) = \frac{q_1(t)}{C_1} + \frac{q_3(t)}{C_{DC}} - \frac{C_d \epsilon q_3(t)^2}{C_{DC}^2} + \frac{C_d^2 \epsilon^2 q_3(t)^3}{C_{DC}^3}$$
$$- \frac{C_{d2} \epsilon^2 q_3(t)^3}{C_{DC}^3} + R_S q_1'(t) + L_1 q_2''(t) + L_3 q_3''(t)$$
$$q_1(t) - q_2(t) - q_3(t) = 0 \quad (3)$$

Furthermore, the coefficients $C_d$ and $C_{d2}$ in (2) represent the first and second-order coefficients relative to the varactor $C(v)$ characteristics, for the chosen biasing condition. For this reason, it is worth pointing out that their values are also functions of the varactor DC bias ($V_{DC}$).

$$V_1 = \frac{X_p + X_o}{C_1} + \frac{Y_p + Y_o}{C_2} + iR_S(X_o\omega_o + X_p\omega_p) + iR_L(Y_o\omega_o + Y_p)\omega_p - L_1(X_o\omega_o^2 + X_p\omega_p^2) - L_2(Y_o\omega_o^2 + Y_p\omega_p^2)$$
$$V_1 = \frac{X_o + X_p}{C_1} + \frac{Z_o + Z_p}{C_{DC}} - \frac{\epsilon C_d Z_o(Z_o + 4Z_p)}{8C_{DC}^2} + \frac{3\epsilon^2 (C_d^2 - C_{d2})(Z_o + Z_p)(Z_o Z_p + Z_o^2 + Z_p^2)}{4C_{DC}^3}$$
$$+ iR_S(X_o\omega_o + X_p\omega_p) - L_1(X_o\omega_o^2 + X_p\omega_p^2) - L_3(Z_o\omega_o^2 + Z_p\omega_p^2)$$
$$X_o + X_p - Y_o - Y_p - Z_o - Z_p = 0 \quad (4)$$

...In (4), $X_p, Y_p, Z_p, X_o, Y_o$ and $Z_o$ represent the generally complex frequency domain components relative to $q_1(t), q_2(t)$ and $q_3(t)$, for both the driving and the sub-harmonic frequencies, respectively, while $\omega_o = 2\pi f_{out}$ and $\omega_p = 2\pi f_{pump}$. It is important to point out that (4) contains terms at both frequencies of interest. Since the validity of (4) must be ensured at both $f_{out}$ and $f_{pump}$, each equation forming it can be divided into two equations, collecting the various terms at these two frequencies. The resulting harmonic balanced system is reported in (5).

$$X_o\left(\frac{-1}{C_1}-iR_S\omega_o+L_1\omega_o^2\right)+Y_o\left(\frac{-1}{C_2}-iR_L\omega_o+L_2\omega_o^2\right)=0$$
$$X_o\left(\frac{-1}{C_1}-iR_S\omega_o+L_1\omega_o^2\right)+Z_o\left(\frac{-1}{C_{DC}}+E+L_3\omega_o^2\right)=0$$
$$X_o - Y_o - Z_o = 0$$
$$\frac{V_1}{2}+X_p\left(\frac{-1}{C_1}-iR_S\omega_p+L_1\omega_p^2\right)+Y_p\left(\frac{-1}{C_2}-iR_L\omega_p+L_2\omega_p^2\right)=0$$
$$\frac{V_1}{2}+X_p\left(\frac{-1}{C_1}-iR_S\omega_p+L_1\omega_p^2\right)+Z_p\left(\frac{-1}{C_{DC}}+F+L_3\omega_p^2\right)=0$$
$$X_p - Y_p - Z_p = 0 \quad (5)$$

In (5), $E$ and $F$ are defined as :

$$E = \frac{\epsilon C_d Z_p}{2C_{DC}^2} - \frac{3\epsilon^2 C_d^2 Z_p^2}{2C_{DC}^3} + \frac{3\epsilon^2 C_{d2} Z_p^2}{2C_{DC}^3}$$
$$F = -\frac{3\epsilon^2 C_d^2 Z_o^2}{2C_{DC}^3} + \frac{3\epsilon^2 C_{d2} Z_o^2}{2C_{DC}^3} \quad (6)$$

From the inspection of (5), it can be observed that all Fourier coefficients ($X_o, Y_o, Z_o, X_p, Y_p$ and $Z_p$) multiply a complex term that includes the static equivalent impedance seen from $N1$, towards one specific branch of the analyzed PFD. As discussed in [30], such an important feature is originated from the dependence of the conversion gain of a modulated capacitor on the impedance that such capacitor sees, from its insertion point and at the different frequencies in the circuit. Consequently, (5) can be used to extract an equivalent transformed two-tones HB-system for PFDs relying on different and generically complex static impedances ($Z_1, Z_2$ and $Z_3$, see Fig. 1). In the following analysis, the value of these impedances, at $f_{out}$ and $f_{pump}$, will be indicated as $Z_1^{(\omega_o)}$, $Z_2^{(\omega_o)}$, $Z_3^{(\omega_o)}$, $Z_1^{(\omega_p)}$, $Z_2^{(\omega_p)}$ and $Z_3^{(\omega_p)}$, respectively. In order to derive the transformed HB-system, two important aspects must be taken into consideration. First, $Z_3$ must include the static impedance ($Z_v$) of the modulated varactor. Also, the excitation voltage ($V_{eq}/2$) that must be used in the transformed HB-system coincides with the applied input voltage ($V_1/2$) only in the typical cases in which the impedance used in the input branch of the PFD is a 1-port network, connected between the input port and $N1$. In contrast, when a 2-port network ($[Z]_{in}$) is used, in the input branch of the PFD, a different excitation voltage must be adopted. Such voltage coincides with the equivalent open-circuit voltage component, at $f_{pump}$, extracted at the output port of $[Z]_{in}$. In this scenario, $V_{eq}$ can be found as $G_v V_1$, where $G_v$ is the open-circuit voltage gain of the equivalent 2-port network formed by the series combination of $R_S$ with $[Z]_{in}$. The resulting HB-system for PFDs using generic $Z_1, Z_2$ and $Z_3$ impedances is reported in (7) and (8). The definition of (7) and (8) allows to extract the response of a PFD as a function of its electrical components and of its modulated capacitance characteristics. To do so, the system in (8) is first solved in terms of $X_p, Y_p$ and $Z_p$ (see (9)), which represent the key components setting the large-signal periodic behavior of PFDs.

$$-iX_o Z_1^{(\omega_o)}\omega_o - iY_o Z_2^{(\omega_o)}\omega_o = 0$$
$$-iX_o Z_1^{(\omega_o)}\omega_o - iZ_o Z_3^{(\omega_o)}\omega_o + EZ_o = 0$$
$$X_o - Y_o - Z_o = 0 \quad (7)$$
$$\frac{V_{eq}}{2} - iX_p Z_1^{(\omega_p)}\omega_p - iY_p Z_2^{(\omega_p)}\omega_p = 0$$
$$\frac{V_{eq}}{2} - iX_p Z_1^{(\omega_p)}\omega_p - iZ_p Z_3^{(\omega_p)}\omega_p + FZ_p = 0$$
$$X_p - Y_p - Z_p = 0 \quad (8)$$

In favor of a simpler analytical treatment, we limit our analysis to the common case in which PFDs use 1-port networks at their input branch, thus making $V_{eq}$ equal to $V_1$.

$$X_p = \frac{3i(C_{d2}-C_d^2)V_1 Z_o^2 \epsilon^2 + 4C_{DC}^3 V_1(Z_2^{(\omega_p)}+Z_3^{(\omega_p)})\omega_o}{6(C_d^2-C_{d2})(Z_1^{(\omega_p)}+Z_2^{(\omega_p)})\omega_o Z_o^2 \epsilon^2 + 8iC_{DC}^3 Z_{eq}^{(\omega_p)}\omega_o^2}$$
$$Y_p = \frac{3i(C_{d2}-C_d^2)V_1 Z_o^2 \epsilon^2 + 4C_{DC}^3 V_1 Z_3^{(\omega_p)}\omega_o}{6(C_d^2-C_{d2})(Z_1^{(\omega_p)}+Z_2^{(\omega_p)})\omega_o Z_o^2 \epsilon^2 + 8iC_{DC}^3 Z_{eq}^{(\omega_p)}\omega_o^2}$$
$$Z_p = \frac{4C_{DC}^3 V_1 Z_2^{(\omega_p)}\omega_o}{6(C_d^2-C_{d2})(Z_1^{(\omega_p)}+Z_2^{(\omega_p)})\omega_o Z_o^2 \epsilon^2 + 8iC_{DC}^3 Z_{eq}^{(\omega_p)}\omega_o^2} \quad (9)$$

In (9), $Z_{eq}^{(\omega_p)}$ is defined as:

$$Z_{eq}^{(\omega_p)} = Z_2^{(\omega_p)} Z_3^{(\omega_p)} + Z_1^{(\omega_p)}\left(Z_2^{(\omega_p)}+Z_3^{(\omega_p)}\right) \quad (10)$$

It is worth emphasizing that (9) include second-order perturbation terms proportional to $Z_o^2$. These terms originate from the expected dynamical changes of the circuit behavior, at $f_{pump}$, in presence of internally generated sub-harmonic oscillations. The existence of these terms is indicative of the ability to use (9) to model the operation of PFDs, both in their dividing and non-dividing operational regimes. However, when we are exclusively interested in determining $P_{th}$, it is possible to simplify (9) by assuming $Z_o$ to be equal to zero. In such case, (9) can be rewritten as:

$$X_p = \frac{-iV_1\left(Z_2^{(\omega_p)}+Z_3^{(\omega_p)}\right)}{2Z_{eq}^{(\omega_p)}\omega_o}$$
$$Y_p = \frac{-iV_1 Z_3^{(\omega_p)}}{2Z_{eq}^{(\omega_p)}\omega_o}$$
$$Z_p = \frac{-iV_1 Z_2^{(\omega_p)}}{2Z_{eq}^{(\omega_p)}\omega_o} \quad (11)$$

Similarly, it is possible to compute $X_o, Y_o$ and $Z_o$ by solving the system of the relative equations in (7). Since we are interested to determine $P_{th}$, (7) can be simplified by observing that low magnitudes of $X_o, Y_o$ and $Z_o$ are still expected when driving the circuit in the close vicinity of its parametric threshold. Consequently, in such scenario of operation, (7) can be rewritten by omitting all second-order perturbation terms proportional to $\epsilon^2$. It is useful to rewrite the resulting set of equations in a matricial representation (12):

$$[A]\begin{bmatrix} X_o \\ Y_o \\ Z_o \end{bmatrix} = 0 \quad (12)$$

where [A] is defined as:

$$\begin{bmatrix} -iZ_1^{(\omega_o)}\omega_o & -iZ_2^{(\omega_o)}\omega_o & 0 \\ -iZ_1^{(\omega_o)}\omega_o & 0 & -iZ_3^{(\omega_o)}\omega_o - \frac{iV_1 C_d Z_2^{\omega_p}}{4C_{DC}^2 Z_{eq}^{(\omega_p)}\omega_o} \\ 1 & -1 & -1 \end{bmatrix} \quad (13)$$

(12) can then be used to identify the voltage amplitude, $V_1$, that triggers the rising of a 2:1 parametric oscillation. To do so, it suffices to find the $V_1$-value ($V_{th}$) that nulls the determinant of the system matrix ([A]). The expression of the so found $V_{th}$-value, as well as the corresponding $P_{th}$, are reported in (14) and (15).

$$V_{th} = \frac{4C_{DC}^2 Z_{eq}^{(\omega_o)} Z_{eq}^{(\omega_p)} \omega_o^2}{C_d \left(Z_1^{(\omega_o)} + Z_2^{(\omega_o)}\right) Z_2^{(\omega_p)}} \quad (14)$$

$$P_{th} = \frac{|V_{th}|^2}{8R_S} \quad (15)$$

In (14) and (15) $Z_{eq}^{(\omega_o)}$ is defined as:

$$Z_{eq}^{(\omega_o)} = Z_2^{(\omega_o)} Z_3^{(\omega_o)} + Z_1^{(\omega_o)} \left(Z_2^{(\omega_o)} + Z_3^{(\omega_o)}\right) \quad (16)$$

As evident from (14), $P_{th}$ is an explicit function of all the impedance values characterizing the operation of a PFD, at both $f_{out}$ and $f_{pump}$. Such impedances significantly shape the stability region of PFDs, thus playing a critical role in their design and performance characteristics. In particular, the inspection of (14) permits to establish general guidelines for the design of PFDs. First, (14) clearly shows that low-capacitance varactors (low $C_{DC}$), with a wide tuning range (large $C_d$), are generally desirable to minimize $P_{th}$. In addition, (14) shows that a quadratic increase of $V_{th}$ is expected at increasing $\omega_o$-values. Such an important feature is mainly due to the increasing challenge of achieving large voltage swings, across variable capacitors, at higher driving frequencies ($\omega_p$). Ultimately, (14) provides an essential guidance in the synthesis of $Z_1, Z_2$ and $Z_3$. In fact, this synthesis can be tackled through conventional linear methods targeting the minimization of (14), even for complex high-frequency board-designs, which are generally sensitive to undesired parasitics. It is important to point out that the ability to select $Z_1, Z_2$ and $Z_3$, without requiring the use of complex time consuming nonlinear algorithms, is essential to enable optimum and reliable performance. It is straightforward to verify that the minimum $P_{th}$ can be attained when four resonant conditions are satisfied. These conditions suggest that:

- the series of $Z_2$ and $Z_3$ has to be designed to series resonate at $f_{out}$ (i.e., $Z_2^{(\omega_o)} + Z_3^{(\omega_o)} \to Re\{Z_2^{(\omega_o)}\} = R'_L$)
- the series of $Z_1$ and $Z_3$ has to be designed to series resonate at $f_{pump}$ (i.e., $Z_1^{(\omega_p)} + Z_3^{(\omega_p)} \to Re\{Z_1^{(\omega_p)}\} = R'_S$)
- $Z_1$ has to be designed to parallel resonate at $f_{out}$ (i.e., $Z_1^{(\omega_o)} \to \infty$)
- $Z_2$ has to be designed to parallel resonate at $f_{pump}$ (i.e., $Z_2^{(\omega_p)} \to \infty$)

In the listed conditions, $R'_S$ and $R'_L$ represent the equivalent resistances observed from $N1$, when looking towards the source and the load, respectively. Even when assuming all components to be loss-less, $R'_S$ and $R'_L$ can be different from $R_S$ and $R_L$. For instance, their value can be made strategically lower through the adoption of impedance transformation stages, as it will be discussed in Section IV. Therefore, when $Z_1, Z_2$ and $Z_3$ are optimally designed, $V_{th}$ becomes:

$$V_{th}^{min} = 4\frac{C_{DC}^2 R'_L R'_S \omega_o^2}{C_d} \quad (17)$$

When the use of a minimum number of lumped components is needed in favor of the highest degree of miniaturization, the optimum synthesis of $Z_1, Z_2$ and $Z_3$ can be tackled through the strategic use of 5 electrical components ($C_1, C_2, L_1, L_2,$ and $L_3$). In particular:

- $Z_1$ can be realized as the parallel combination of an inductor ($L_1$) and a capacitor ($C_1$), whose resonance frequency matches the output frequency at which the minimum $P_{th}$ is desired;
- $Z_2$ can be realized as the parallel combination of an inductor ($L_2$) and a capacitor ($C_2$), whose resonance frequency is equivalent to twice the output frequency at which the minimum $P_{th}$ is desired;
- $Z_3$ includes the static portion ($C_{DC}$) of the varactor electrical characteristics; it can be realized by adding an inductor ($L_3$), whose value is directly related to both $Z_2^{(\omega_o)}$ and $Z_1^{(\omega_p)}$.

Note that $P_{th}$ is a function of all the abovementioned parameters. Among them, $C_{DC}$ depends on the available varactor technology, whereas $L_1, C_1, L_2$ and $C_2$ depend heavily on the chosen value of $L_3$. Consequently, it is convenient to search for the optimal values of $L_1, C_1, L_2$ and $C_2$ that satisfy the listed resonant conditions in terms of $C_{DC}$ and $L_3$(see (18)).

$$C_1 = \frac{4C_3}{3\left(-1 + 16L_3 C_3 f_{out}^2 \pi^2\right)}$$
$$C_2 = -\frac{4C_3}{3\left(-1 + 4L_3 C_3 f_{out}^2 \pi^2\right)}$$
$$L_1 = \frac{3\left(-1 + 16L_3 C_3 f_{out}^2 \pi^2\right)}{16 C_3 f_{out}^2 \pi^2}$$
$$L_2 = -\frac{3\left(-1 + 4L_3 C_3 f_{out}^2 \pi^2\right)}{16 C_3 f_{out}^2 \pi^2} \quad (18)$$

It is important to point out that, despite of the fact that a low $C_{DC}$-value would be required to ensure the minimum $P_{th}$, the use of ultra-low capacitance varactors is not practical as it often leads to sub-optimal performance. This important feature is mostly determined by the limited $Q$ of available inductors. This limitation constrains $L_1, L_2$ and $L_3$ not to exceed certain values, in order to prevent undesired and significant increases of $R'_L$ and $R'_S$. As a case study, it is now instructive to extract $V_{th}$ through (14) for a generic simplified PFD-design, formed by loss-less inductors and capacitors. The resulting value can then be compared with the one numerically found through TD-based methods, in order to confirm the validity of the reported analytical approach. The chosen device, a 200:100MHz PFD, as well as the adopted $L_3, C_{DC}, C_d$, and $C_{d2}$ values along with the corresponding optimal $L_1, L_2, C_1$ and $C_2$ values, are reported in Fig. 2.

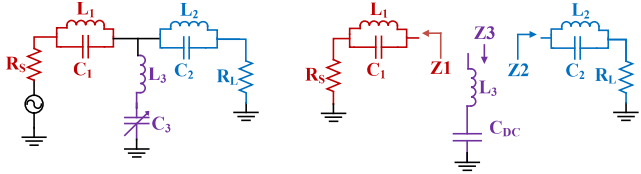

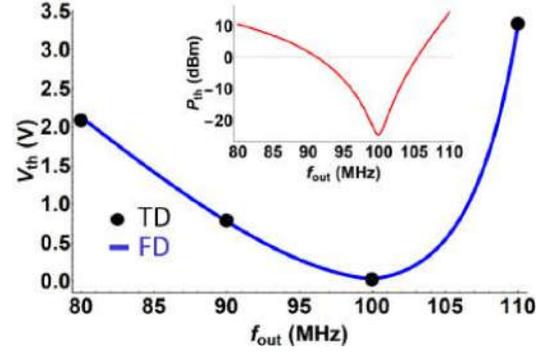

Fig. 2. Left) Schematic view of the 200:100MHz PFD used as a case study in this work. The values of the adopted circuit components are: $L_1 = 382.5\ nH$, $L_2 = 742.5\ nH$, $C_1 = 6.6\ pF$, $C_2 = 0.85\ pF$, $L_3 = 500\ nH$, $C_{DC} = 1.7\ pF$, $C_d = -0.3$, $C_{d2} = 0.02$; Right) Schematic representation for $Z_1$, $Z_2$ and $Z_3$ for the same PFD.

Fig. 3. In black), $V_{th}$-value points that were numerically determined through lengthy time-domain integration methods applied to the PFD shown in Fig. 2; in blue), calculated $V_{th}$-values from (14) for the same circuit; in red), $P_{th}$-values vs. $f_{out}$ corresponding to the $V_{th}$-values plotted in blue.

The estimated $V_{th}$-values based on (14) are shown in Fig. 3 together with some corresponding results, for different $f_{out}$-values (keeping $f_{pump} = 2f_{out}$), obtained through TD-methods. Clearly, the $V_{th}$-values derived through numerical TD-methods match very closely those analytically predicted through (14), thus demonstrating the validity of the analytical findings. The $P_{th}$-values corresponding to the $V_{th}$-values found through (14) are also shown in Fig. 3. As evident, when neglecting the ohmic losses introduced by each adopted element and when properly selecting the different components forming its network, the investigated PFD can exhibit a $P_{th}$-value (for $f_{out}$=100MHz) that is lower than -24 dBm, corresponding to a $V_{th}$ of 0.038V. To further verify the substantial and desired change in the dynamical behavior of the analyzed PFD, for $V_1$ being slightly higher or slightly lower than $V_{th}$, we report the phase portraits [37] (Fig. 4) relative to $q_3(t)$ and derived, through TD-methods, when assuming $f_{out}$ equal to 100MHz (and, consequently, $f_{pump}$ equal to 200MHz) for $V_1$ equal to 0.037V and 0.039V, respectively. As evident, substantially different behaviors characterize the operation of the analyzed PFD for the two investigated $V_1$-values. In particular, for $V_1$ equal to 0.037 V the portrait of $q_3(t)$ shows the existence of a limit-cycle. This cycle maps the evolution of $q_3(t)$ and $q'_3(t)$ as time evolves from the origin of the reference system (t=0) and zero pre-stored charge exists in the different capacitors. In contrast, for $V_1$ equal to 0.039 V (thus being higher than the expected $V_{th}$-value), the portrait exhibits a substantially different behavior. In fact, once the PFD reaches its steady-state periodic operational regime (see the red-lines in Fig. 4), the portrait exhibits twice the period it exhibits in the former case. Such a unique feature maps (in time-domain) the origin of a period-doubling mechanism that marks the existence of a sub-harmonic oscillation in the circuit. The phenomenon of period-doubling can also be directly observed by extracting the Poincaré map [38] (PM, see Fig. 5). This tracks the radius of the limit-cycle of $q_3(t)$ ( $r = \sqrt{(q_3(t+nT_s)^2 + \left(\frac{q'_3(t+nT)}{\omega_p}\right)^2}$, where $n$ is an integer number) vs. $V_1$, for consecutive returns (incremental $n$-values) on the map. This can be done by sampling the dynamical steady-state large-signal TD response of the analyzed PFD with a sampling rate that is equal to the excitation frequency, while assuming a continuously increasing driving voltage. As evident in Fig. 5, for $V_1 < V_{th}$ the PM shows a continuous trend with respect to $V_1$. Such a unique feature is a clear indication that $q_3(t)$ has a period that is equal to $1/f_{pump}$. In contrast, for $V_1 \geq V_{th}$, the PM exhibits two separate lines. These indicate the values of $r$, at the end of consecutive sampling periods and for continuously increasing $V_1$-levels. The existence of two consecutive lines in PMs is a clear indication of the different $q_3(t)$ values that the system exhibits (above threshold) at the end of consecutive sampling periods. This dynamical feature is also often used to numerically identify, through TD-methods, the presence of period-doubling regimes in nonlinear dynamical systems.

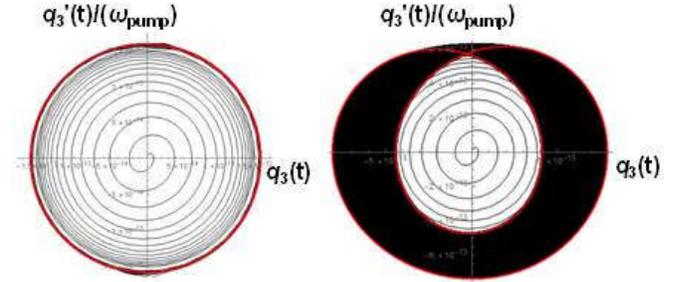

Fig. 4. Left) Phase portrait of the charge ($q_3(t)$) flowing in the varactor in the analyzed PFD design for $V_1 = 0.037\ V$ (thus below $V_{th}$). This plot was computed after assuming the varactor to be completely discharged at the moment ($t = 0$) in which the input voltage, $V_1$, is applied to the circuit. In red, the typical limit-cycle describing the dynamical behavior of the analyzed PFD, after reaching its steady-state periodic characteristics, is also reported. Right) Phase portrait of $q_3(t)$ for $V_1 = 0.039\ V$ (thus above $V_{th}$). This plot was computed after assuming the varactor to be completely discharged at the moment ($t = 0$) in which the input voltage, $V_1$, is applied to the circuit. After a limited number of excitation cycles, the portrait exhibits a period twice the excitation period. Consequently, once the PFD reaches its steady-state periodical regime, the trajectory described by the system (see the red-line) assumes a different shape from conventional limit-cycles.

Instead, in order to confirm that the analytically derived $P_{th}$-values closely match the ones found with a commercial circuit simulator, through the pAG-technique (Section II), we compared the distributions of $P_{th}$ vs. $L_3$ extracted, both analytically and through the application of the pAG-technique, for the circuit in Fig. 2. As can be observed from Fig. 6, the distribution of $P_{th}$ vs. $L_3$, derived through the pAG-technique, agrees with the one analytically calculated using (15). This further demonstrates the validity of our analytical findings. It is worth mentioning that, in order to extract the simulated data points shown in Fig. 6 (see red points), a $P_{in}$-sweep of 6000 steps had to be configured for the HB-simulator to converge to the desired non-trivial solution. This constraint can lead to design times, for optimized PFDs, that can exceed days. Ultimately, we analytically studied the dependence of $P_{th}$ on the $Q$-value that can be exhibited by practical inductors forming the circuit shown in Fig. 2 (see inset in Fig. 6). In order to do so, for simplicity, we assumed that all the inductors were exhibiting the same $Q$. Interestingly, our investigation revealed that ultra-low $P_{th}$-values can indeed be attained, even when using inductors with $Q$-values that are only around 50.

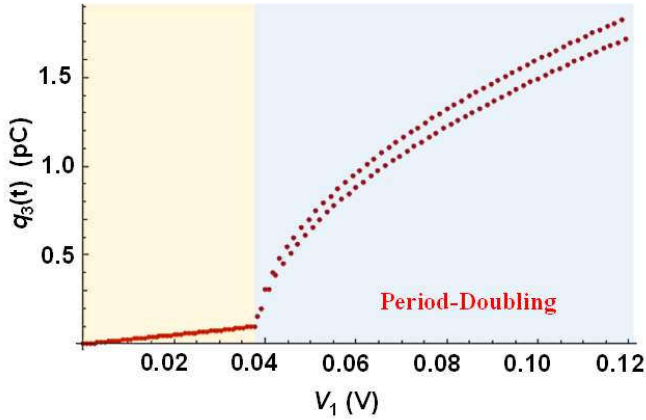

Fig. 5. Poincaré map relative to the charge $q_3(t)$ vs. the peak-value ($V_1$) of the input voltage $V_1(t)$. As evident, for $V_1$-values that exceed the same $V_{th}$-value that was found through our HB-based approach, the analyzed PFD undergoes a change in its dynamical characteristics that results into the activation of a period-doubling mechanism.

### B. Evaluating the Response of PFDs for $P_{in} > P_{th}$

In the previous section, a closed-form expression was found to evaluate the threshold voltage and power ($V_{th}$ and $P_{th}$) activating a sub-harmonic oscillation in varactor-based PFDs. In this section, the procedure to analytically estimate the complete response of PFDs below and above the parametric threshold is discussed. The assessment of the PFD response, after the occurrence of a bifurcation, requires solving the systems in (7) and (8) without neglecting any second-order perturbation term proportional to $\epsilon^2$. This increased complexity renders the solution of these systems only computable through numerical methods. However, some important features can still be identified. In fact, three sets of $X_o, Y_o$ and $Z_o$ values are found to be potential solutions for the system in (7) and (8). One set (S1) is representative of the trivial solution ($X_o = Y_o = Z_o = 0$), thus describing the evolution of the PFD when no input signal exists in the circuit at $f_{out}$, and when assuming $V_1 << V_{th}$. Another set (S3) shows quasi-uniform and not-nulled distributions for $X_o, Y_o$ and $Z_o$ vs. the magnitude of $V_1$. Finally, the last solution set (S2) corresponds to more complex distributions for $X_o, Y_o$ and $Z_o$ vs. $V_1$, which exhibit nulled values for $V_1$ approaching $V_{th}$. The amplitudes of the voltage components across $R_L$ ($V_{out}$), at $f_{out}$, for solutions S1, S2 and S3 were analytically determined and plotted versus the input power ($P_{in}$), at $f_{pump}$, for the investigated PFD in Fig. 2. Fig. 7 displays the extracted trends of $V_{out}$ vs. $P_{in}$ for S1, S2 and S3.

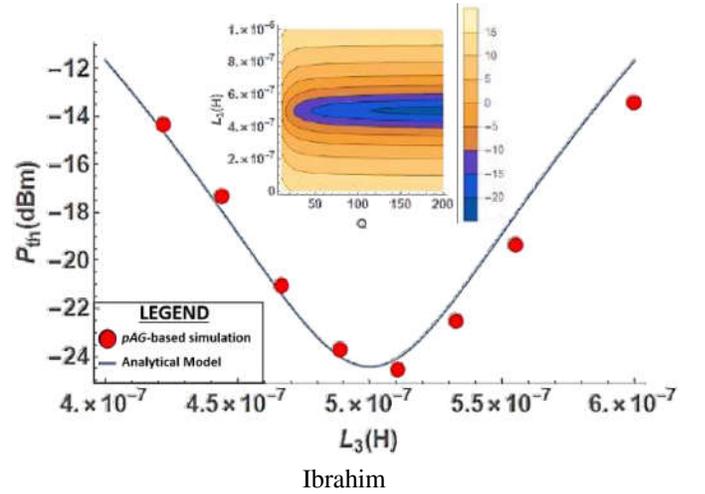

Ibrahim

Fig. 6. Analytical trend (in blue) of $P_{th}$ vs. $L_3$ (Fig. 1) and simulated values, through the pAG-technique (in red) and for a limited set of $L_3$-values, relative to the PFD described in Fig. 3. It is important to point out that the HB simulator, adopted to estimate $P_{th}$ through the pAG-technique, has been configured to include 28 harmonics of the output frequency ($f_{out}$) in favor of a more accurate prediction of the PFD response, above threshold. In the inset, a contour-plot is reported mapping, simultaneously, the impact on $P_{th}$ of $L_3$ and of the quality factor (Q) exhibited, for simplicity, by all the inductors used by the same circuit.

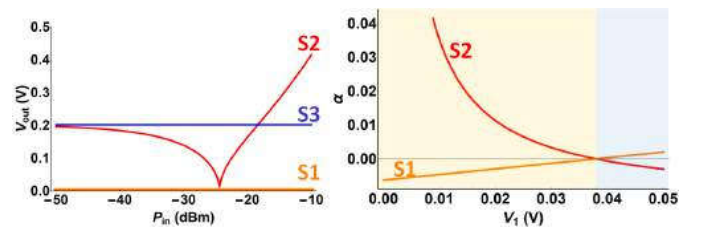

Fig. 7. Left) solution amplitudes of the possible voltage ($V_{out}$) components (at $f_{out}$) across the load resistance $R_L$ of the PFD shown in Fig. 2. Note that three different $V_{out}$ distributions (relating to the solution sets S1, S2 and S3) are solution of (7). Right) Real part of the eigenvalue ($\alpha$) of the Jacobian matrices relating to (7) and linearized around the $X_o, Y_o$ and $Z_o$ values that define S1 and S2. The $\alpha$-value of S3 is constant and positive for all $V_1$-values, thus being a clear indication that S3 is not a viable solution for the system. For this reason, its distribution with respect to $V_1$ has not been included here.

Due to the existence of three possible steady-state periodic solutions (S1, S2 and S3), the complete response of the analyzed PFD can only be determined by evaluating the stability of each solution, while varying the magnitude of the applied input signal. In order to do so, the matrix-A (12), linearized around each solution, can be used to describe the evolution of the system in presence of small perturbations acting on the steady-state amplitudes of $X_o, Y_o$ and $Z_o$ relative to the same solution. It is worth pointing out that the resulting linearized matrix represents a Jacobian matrix ([J]) that provides the means to investigate the stability of any possible solution at $f_{out}$. This can be done by looking at the sign of the purely real eigenvalue ($\lambda$) of [J] (see Fig. 7). In particular, by looking at the real part ($\alpha$) of $\lambda$, it is straightforward to realize that S3 corresponds to a positive and constant $\alpha$-value for any $V_1$-value, thus representing a known unstable point for the system. In contrast, for both S1 and S2, the sign of $\alpha$ changes when $V_1$ approaches $V_{th}$. In particular, S1 represents the only stable solution for $V_1 < V_{th}$, whereas S2 represents the only stable solution for $V_1 \geq V_{th}$. In other words, the trivial solution is stable for $V_1 < V_{th}$, whereas the dividing solution is stable for $V_1 \geq V_{th}$. It is also worth pointing out the fact that S1 and S2 flip their stability at the same $V_1$-value, suggesting that the rising of the sub-harmonic oscillation occurs through a supercritical bifurcation. For this reason, no abrupt jump is expected in the PFD frequency response, as its operation involves the transition from one operational regime to the other. After determining the stability of S1 and S2, it is easy to estimate the output power of the PFD in Fig. 2 when this is driven at $f_{pump}$=200MHz and when $P_{in}$ is progressively increased to activate the division process in the circuit. In Fig. 8, we report the PFD output power ($P_{out}$, delivered to $R_L$) vs. $P_{in}$ for the PFD shown in Fig. 2.

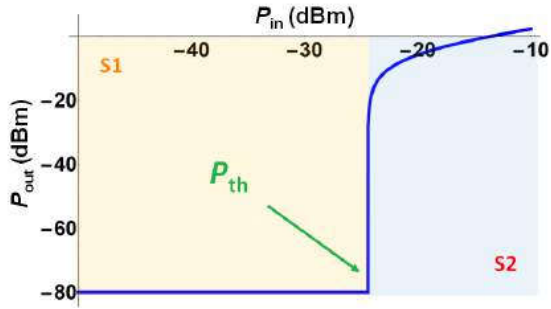

Fig. 8. Extracted output power ($P_{out}$) characteristics of the analyzed PFD, showing a $P_{th}$-value that approaches $-25$ $dBm$. The plot was generated with the assumption of a white noise power spectral density in the circuit, resulting into a $-80$ $dBm$ of power at $f_{out}$ for $P_{in} < P_{th}$. (when the trivial solution is only stable solution for the circuit).

## IV. A LUMPED REALIZATION OF A 200:100 MHZ PFD

In the previous sections we derived a closed-form expression for $P_{th}$ and described a method to analytically estimate the steady-state output response of PFDs, for $P_{in}$-values that are lower or higher than $P_{th}$. In order to experimentally verify the validity of our findings, we designed a 200:100MHz PFD, using lumped components available on the shelf. The PFD was assembled on a PCB made of FR4, and its performance was characterized using conventional RF measurement equipment. A detailed description of the adopted design flow, as well as the analysis of our measured results, is discussed in the remainder of this section.

### A. Design of A 200:100 MHz PFD

The design of the reported PFD targeted the minimization of $P_{th}$, at a chosen $f_{out}$-value of interest. 100MHz was chosen here as the desired output frequency. Fig. 9 displays a schematic representation of the PFD architecture that we selected for the experimental validation. This PFD design relies on the five components ($L_1, L_2, C_1, C_2$ and $L_3$, see (18)) used in the simplified PFD circuit in Fig. 2. However, two additional components were added, a capacitor and an inductor. These components, labeled as $C_{matching}$ and $L_{matching}$, were chosen so as to form the equivalent lumped representation of a quarter-wave transformation stage, at $f_{out}$. It is important to point out that the adoption of this stage is key to lower the $P_{th}$-value that can be attained through the only use of the other five components. In fact, the adoption of this stage reduces the impact of the output load ($R_L$) on the stability of the PFD, by converting $R_L$ to an impedance ($Z_{Trans}$) whose value, at $f_{out}$, is real and lower than 1$\Omega$. In other words, the use of the transformation stage permits to reduce $R_L^{'}$, thus minimizing $V_{th}^{min}$ and, consequently, $P_{th}$ (see (15) and (17)). Moreover, the adoption of a transformation network, relying on a series capacitor, also allows to use $C_{matching}$ as a DC-blocker. The selected values for $C_{matching}$ (227 pF) and $L_{matching}$ (11.3 nH) were chosen so that the lowest $Z_{Trans}$ could be attained, when considering the typical $Q$ values exhibited by available inductors on the shelf. Because of the more dispersive behavior of $Z_2$, with respect to the one of the simplified circuit in Fig. 2, the resonant conditions that must be satisfied in order to minimize $P_{th}$ lead to slightly different expressions for the optimal values of $L_1, L_2, C_1$ and $C_2$, vs. $L_3$ and $C_3$. These expressions are rather cumbersome, which is why they are being omitted here. Furthermore, the component values significantly depend on the maximum $Q$ that can be exhibited by $L_1$ and $L_2$ in practice. Therefore, after finding a commercial available hyper-abrupt varactor (model: Skyworks SMV1405), characterized by $C_{DC}$ values ranging from 1 pF to 10 pF and capable to exhibit high $C_d$ values (see (2)), we analytically studied the distribution of $P_{th}$ (see (15)) vs. $L_3$ and $C_{DC}$. In order to do so, the $C_d$ value exhibited by the selected varactor was expressed in terms of $C_{DC}$. This simplification made $C_{DC}$ the only required varactor parameter to extract $P_{th}$. Moreover, the derived $P_{th}$ distribution was found after selecting the $L_1, L_2, C_1$ and $C_2$ values that satisfy the resonant conditions discussed in Section III-A, for each analyzed set of $L_3$ and $C_{DC}$ values. In Fig. 10 and in Fig. 11 we report the computed trend and the contour plot of $P_{th}$ vs. $L_3$ and $C_{DC}$ when assuming the $Q$ of $L_1$, $L_2$ and $L_3$ to be 50. This value corresponds to the best quality factor that we could find, for inductors that are in the same range than those required to optimally design a 200:100MHz PFD, relying on ideal loss-less components to work (see Fig. 2). As evident, a monotonically decreasing $P_{th}$ is attained as $L_3$ is increased. However, the adoption of $L_3$ values larger than 800nH would require $C_2$ to be lower than 0.5pF. This design constraint would expose any PFD built on a PCB to the risk of exhibiting performance that are too sensitive to un-modelled variations of the actual $C_2$ value. Also, because of the limited availability of surface-mount commercial inductors, simultaneously showing large inductance, high-$Q$s (exceeding 50) and a self-resonance frequency higher than the maximum frequency of interest (200 MHz), the use of excessively large $L_3$ values is also not practical. Based on these limitations, we selected the $C_{DC}$ value (1.7 pF) that minimizes $P_{th}$, given the largest suitable $L_3$-value that we could find (500 nH). By looking at the $C(V)$ characteristics of the chosen varactor, this optimal $C_{DC}$ value allows to determine the corresponding DC-voltage (1.6 V) that must be used to bias the varactor, in the actual PFD circuit. Also, after selecting $L_3$, we looked at the sensitivity of the optimal $C_{DC}$ value, as we vary the $Q$ exhibited by $L_1$, $L_2$ and $L_3$ ranging from 10 to 50. As evident from Fig. 12, we found the optimal $C_{DC}$ value to be only slightly dependent on the $Q$ of the adopted inductors, thus being almost immune to non-idealities that often make commercial inductors exhibit different $Q$s from their nominal values. In summary, the values that we selected for the experimental demonstration are 382.5 nH, 742.5 nH, 6.6 pF, 0.85 pF and 500 nH for $L_1, L_2, C_1, C_2$ and $L_3$, respectively. Based on these values, we searched for the commercial components with the closest nominal behavior to the desired ones. Then, we assembled a distributed model of the the board layout, using microstrip components. After building this model, we minimized the impact of the board layout on $P_{th}$. In order to do so, we extracted the corresponding values of $Z_1^{(\omega_o)}$, $Z_2^{(\omega_o)}$, $Z_3^{(\omega_o)}$, $Z_1^{(\omega_p)}$, $Z_2^{(\omega_p)}$ and $Z_3^{(\omega_p)}$. These impedances were then used to generate a layout optimization flow, targeting the minimum $P_{th}$, at $f_{out}$=100 MHz. During this optimization step, we also considered the available S-parameters for the lumped components that we selected. We report, in Fig. 13 (see green line), the analytically derived distribution of $P_{th}$ vs. $f_{out}$, extracted through (15), after determining the best layout geometry and under the assumption that the input frequency is always twice the $f_{out}$ value. The same distribution was also evaluated using the pAG technique by replacing the optimized distributed model for the board with its actual electromagnetic simulated RF model (see red points in Fig. 13). It can be seen that both simulation approaches exhibit closely matching trends and predicted minimum $P_{th}$ values lower than -12.5 dBm, at $f_{out}$ equal to 100 MHz. However, for $f_{out}$ lower than 97 MHz, the HB simulation based on the pAG technique and using the electromagnetic model of the board was not able to converge to the dividing solution. Thus, $P_{th}$ values could only be extracted for $f_{out}$ values above 97MHz.

### B. Measured Results

The designed PFD was built on a PCB made of FR4 (see inset of Fig. 14). An external bias-T (model Inmet 8800SMF3-06) was used to simultaneously drive the PFD input port with an RF signal and with a DC voltage (1.6 V) required to properly bias the varactor. The output performance of the PFD, terminated on a 50Ω resistive load, was characterized using conventional RF bench-top measurement equipment. In particular, to extract the PFD output power ($P_{out}$) at the desired output frequency (100 MHz), we used two synced vector network analyzers (VNAs). One network analyzer (Keysight PNA N5221A) was set up to produce the pump signal at 200MHz and to generate a slow $P_{in}$ sweep ranging from -25 dBm to 0 dBm. The other VNA (Keysight ENA E5071C) was used to track the received power at 100MHz. The measured distribution of $P_{out}$ vs. $P_{in}$ (Fig. 14) closely follows the predicted distribution found through the pAG technique. It is worth mentioning that a much higher $P_{th}$ would have been attained without the use of a transformation stage (see the simulated green trend in Fig. 14). To visualize the output response of the PFD, after the activation of the division process, the measured TD-waveform of its output voltage, across $R_L$, is shown in Fig. 15, for a $P_{in}$-value (-2dBm) exceeding $P_{th}$. Under this operating condition, the presence of an output signal, with a strong frequency component at 100 MHz, can be easily observed. Ultimately, the measured distribution of $P_{th}$ vs. $f_{out}$, for $f_{out}$ values ranging from 90 MHz to 110 MHz and using a $f_{pump}$ value that is always twice $f_{out}$, is reported in Fig. 13. As evident, this distribution agrees with our predictions, found through the described analytical method and through the use of the pAG-technique.

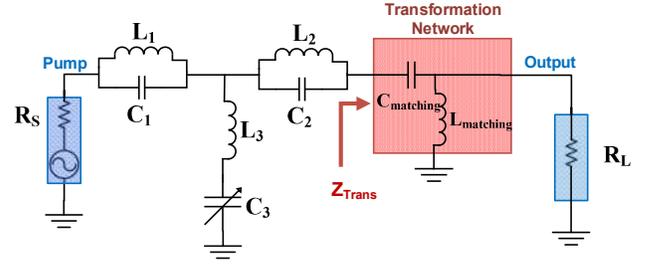

Fig. 9. Schematic representation of the example PFD built in this work. The circuit is designed to be driven by a 50 Ω generator, and to be attached (at its output) to a 50 Ω output load ($R_L$).

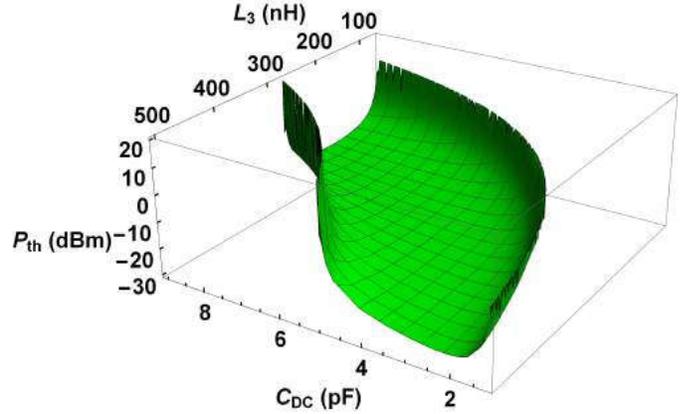

Fig. 10. A 3D plot mapping the distribution of $P_{th}$ vs. $L_3$ and $C_{DC}$, where a $Q$ equal to 50 was assumed for all inductors in the circuit.

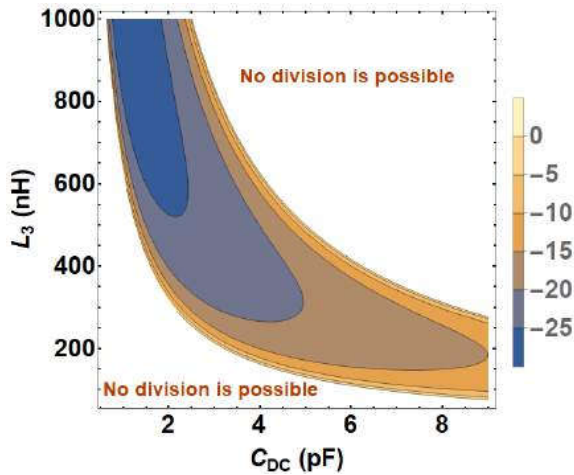

Fig. 11. Contour plot mapping $P_{th}$ vs. $L_3$ and $C_{DC}$, assuming that the $Q$ of all inductors is 50. The locus of these two parameters where no division is possible is indicated.

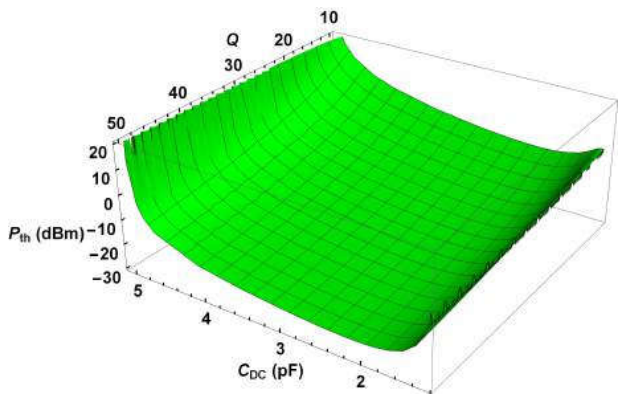

Fig. 12. A 3D plot mapping the distribution of $P_{th}$ vs. $C_{DC}$ and $Q$ for the selected $L_3$ value of 500nH.

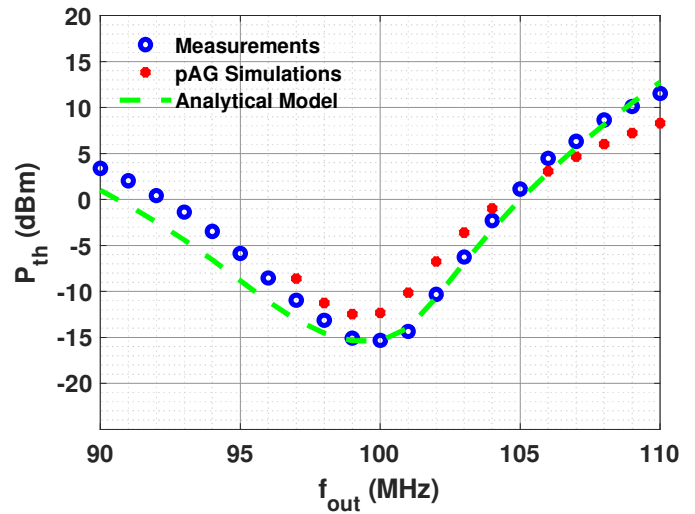

Fig. 13. Simulated distributions of $P_{th}$ vs. $f_{out}$ for the PFD from this work: In green), simulated values extracted through the reported analytical method using a distributed model for the layout. In red), simulated values extracted through the pAG technique. Here, the $P_{th}$ values for $f_{out}$ lower than 97 MHz could only be obtained because of the lack of convergence of the pAG technique with electromagnetic simulations. In blue), measured distribution of $P_{th}$ vs. $f_{out}$ for the same frequencies used during simulations. In order to extract both the simulated and the measured data, $f_{pump}$ was kept equal to $2f_{out}$ for all investigated $f_{out}$ values.

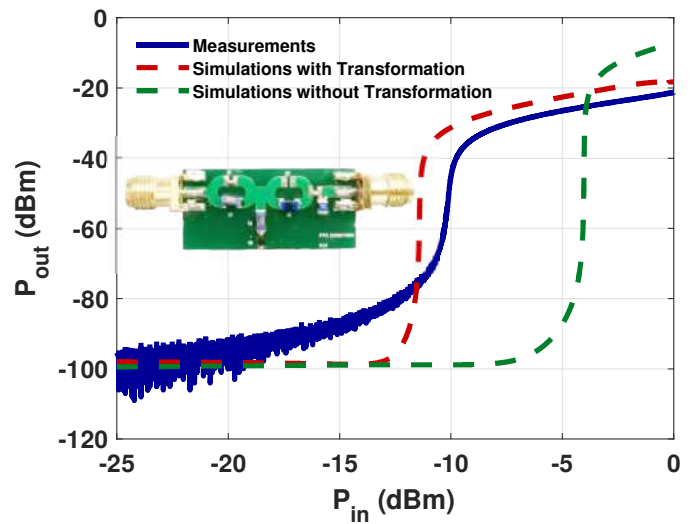

Fig. 14. Measured (in blue) and simulated (in red, using the pAG technique) distributions of $P_{out}$ vs. $P_{in}$ for the PCB from this work. For comparison, the simulated distribution of $P_{out}$ of the same PFD without transformation stage is shown in green. The inset displays a photo of the fabricated PFD.

## V. CONCLUSION

In this work, a new systematic synthesis approach is discussed to enable the design of varactor-based parametric frequency dividers, exhibiting ultra-low power threshold ($P_{th}$). In particular, it is analytically shown, for the first time, that the $P_{th}$ value exhibited by PFDs can be written as a closed-form explicit function of the impedances seen by the variable capacitor, at the main input and output frequencies of operation. This unique feature permits to create optimum PFD designs, without relying on the time-consuming and memory-intensive simulation approaches that are currently available, but only through conventional design and optimization techniques that are frequently used in linear circuits. Thanks to the development of the reported analytical framework, we formulate a new optimal design criteria for PFDs requiring to exhibit ultra-low $P_{th}$-values. In order to experimentally validate our analytical findings, a 200:100MHz PFD, relying on commercially available lumped components, was designed and assembled on a printed-circuit-board (PCB). Thanks to its engineered design and despite the relatively low $Q$ exhibited by its inductors, the fabricated PFD exhibits a record-low $P_{th}$-value, in pair of -15dBm. The new design approach, presented in this work, opens exciting scenarios for the development of even other parametric components. In particular, the capability to obtain ultra-low threshold PFDs will facilitate the future chip-scale development of parametric filters (PFILs), thus enabling their use to reduce the jitter exhibited by currently available frequency synthesizers, in low-power RF transceivers.

## ACKNOWLEDGMENT

This work has been funded by the National Science Foundation (NSF), under the awarded grant: 1854573.

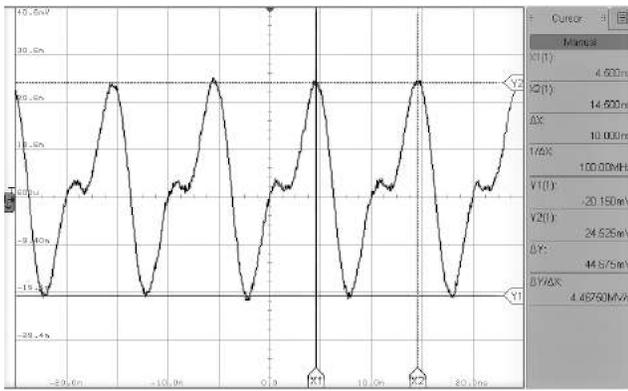

Fig. 15. Measured waveform, relative to the output voltage of the PFD, extracted from a high frequency oscilloscope (Keysight DSOX6004A) when $f_{out}$ was set to 100 MHz (i.e. $f_{pump}$=200 MHz) and $P_{in}$ was chosen to be -2 dBm, thus much higher than $P_{th}$

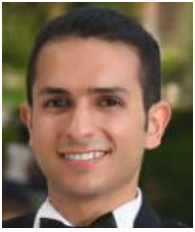
**Hussein M. E. Hussein** (S'20) received his B.S. and M.Sc. in electrical engineering at Cairo University, Giza, Egypt, in 2013 and 2017, respectively. He is currently pursuing the Ph.D. degree with the Electrical and Computer Engineering department at Northeastern University, Boston, MA, USA. He is currently working on the development of parametric phase noise reduction techniques, for RF systems, based on nonlinear devices and circuits.

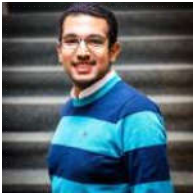
**Mahmoud A. A. Ibrahim** (S'13) received the B.Sc. (Hons.) and M.Sc. degrees in electrical engineering from the Electronics and Electrical Communications Engineering Department, Cairo University, Giza, Egypt, in 2013 and 2015, respectively. He is currently pursuing the Ph.D. degree in electrical engineering with Northeastern University, Boston, MA, USA.
From 2013 to 2015, he was a Teaching and Research Assistant with Cairo University. In the summer of 2018, he joined the PLL Team at Qualcomm, San Diego, CA, USA, as an Analog-Mixed Signal Design Intern; where he was involved in the research and design of ultra-low power multi-Giga Hertz LC-oscillators in deep-submicron Fin-FET technologies. Since 2016, he has been a Graduate Research and a Teaching Assistant with Northeastern University. He is also working on the design of ultra-low power transceivers for biomedical applications. His research interests include integrated analog, mixed-signal, RF circuits for low-power wireless transceivers, and power management integrated circuits.

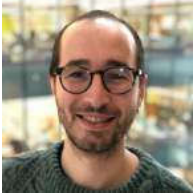
**Giuseppe Michetti** is a Ph.D. Student at the Electrical and Computer Engineering department at Northeastern University, Boston (USA). He received his B.S. and M.Sc. at Politecnico di Milano (Italy) in 2014 and 2018, respectively. His research is based MEMS applications for microwave circuits for mobile platforms. Currently he is working on nonlinear and time-variant models and prototypes for RF circuits based on piezoelectric RF MEMS technologies for novel generation of RF front-ends.

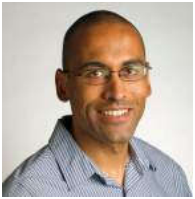
**Marvin Onabajo** (S'01–M'10–SM'14) is an Associate Professor in the Electrical and Computer Engineering Department at Northeastern University. He received a B.S. degree (summa cum laude) in Electrical Engineering from The University of Texas at Arlington in 2003, as well as the M.S. and Ph.D. degrees in Electrical Engineering from Texas A&M University in 2007 and 2011, respectively. From 2004 to 2005, he was Electrical Test/Product Engineer at Intel Corp. in Hillsboro, Oregon. He joined the Analog and Mixed-Signal Center at Texas A&M University in 2005, where he was engaged in research projects involving analog built-in testing, data converters, and on-chip temperature sensors for thermal monitoring. In the spring 2011 semester, he worked as a Design Engineering Intern in the Broadband RF/Tuner Development group at Broadcom Corp. in Irvine, California. Marvin Onabajo has been at Northeastern University since the Fall 2011 semester. His research areas are analog/RF integrated circuit design, on-chip built-in testing and calibration, mixed-signal integrated circuits for medical applications, data converters, and on-chip sensors for thermal monitoring. He currently serves as Associate Editor on the editorial boards of the IEEE Transactions on Circuits and Systems I (TCAS-I, 2016-2017, 2018-2019, and 2020-2021 terms) and of the IEEE Circuits and Systems Magazine (2016-2017, 2018-2019, and 2020-2021 terms). During the 2014-2015 term, he was on the editorial board of the IEEE Transactions on Circuits and Systems II (TCAS-II). He received a 2015 CAREER Award from the National Science Foundation, a 2017 Young Investigator Program Award from the Army Research Office (ARO), and the 2015 Martin Essigman Outstanding Teaching Award from the College of Engineering at Northeastern University.

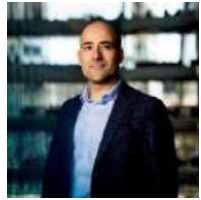
**Matteo Rinaldi** is an Associate Professor in the Electrical and Computer Engineering department at Northeastern University and the Director of Northeastern SMART a university research center that, by fostering partnership between university, industry and government stakeholders, aims to conceive and pilot disruptive technological innovation in devices and systems capable of addressing fundamental technology gaps in several fields including the Internet of Things (IoT), 5G, Quantum Engineering, Digital Agriculture, Robotics and Healthcare. Dr. Rinaldi received his Ph.D. degree in Electrical and Systems Engineering from the University of Pennsylvania in December 2010. He worked as a Postdoctoral Researcher at the University of Pennsylvania in 2011 and he joined the Electrical and Computer Engineering department at Northeastern University as an Assistant Professor in January 2012. Dr. Rinaldi's group has been actively working on experimental research topics and practical applications to ultra-low power MEMS/NEMS sensors (infrared, magnetic, chemical and biological), plasmonic micro and nano electromechanical devices, medical micro systems and implantable micro devices for intra-body networks, reconfigurable radio frequency devices and systems, phase change material switches, 2D material enabled micro and nano mechanical devices. The research in Dr. Rinaldi's group is supported by several Federal grants (including DARPA, ARPA-E, NSF, DHS), the Bill and Melinda Gates Foundation and the Keck Foundation with funding of $14+M since 2012. Dr. Rinaldi has co-authored more than 140 publications in the aforementioned research areas and also holds 10 patents and more than 10 device patent applications in the field of MEMS/NEMS. Dr. Rinaldi was the recipient of the IEEE Sensors Council Early Career Award in 2015, the NSF CAREER Award in 2014 and the DARPA Young Faculty Award class of 2012. He received the Best Student Paper Award at the 2009, 2011, 2015 (with his student) and 2017 (with his student) IEEE International Frequency Control Symposiums; the Outstanding Paper Award at the 18th International Conference on Solid-State Sensors, Actuators and Microsystems, Transducers 2015 (with his student) and the Outstanding Paper Award at the 32nd IEEE International Conference on Micro Electro Mechanical Systems, MEMS 2019 (with his student). Prof. Rinaldi is the founder and CEO of Zepsor Technologies, a start-up company that aims to bring to market zero standby power sensors for various internet of things applications including distributed wireless fire monitoring systems, battery-less infrared sensor tags for occupancy sensing and distributed wireless monitoring systems of plant health parameters for digital agriculture. Prof. Rinaldi is also the owner of Smart MicroTech Consulting LLC, a company that routinely provides consulting services to government agencies, large companies and startups in the broad areas of Micro and Nano Technologies, Internet of Things, Wireless Communication devices and systems, Radio Frequency Devices and Systems and Sensors.

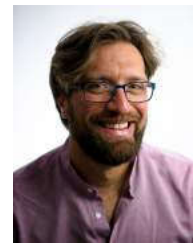
**Cristian Cassella** is an Assistant Professor in the Electrical and Computer Engineering department at Northeastern University, Boston (USA). He received his B.S.E. and M.Sc., with honors, at University of Rome – Torvergata in 2006 and 2009, respectively. In 2011, he was a visiting scholar at University of Pennsylvania. In 2012 he entered a Ph.D. program at Carnegie Mellon University which he completed in 2015. In 2015 he was a Postdoctoral Research Associate at Northeastern University. In 2016, he became Associate Research Scientist. He is author of 60 publications in peer-reviewed journals and conference proceedings. Two of his peer-reviewed journal papers published on the IEEE Journal of MicroElectroMechanical systems (JMEMS) were selected as papers of excellent quality (JMEMS RightNowPapers), hence being released as open-access. One of his journal papers was chosen as the cover for the Nature Nanotechnology October 2017 issue. He won the best paper award at the IEEE International Frequency Control Symposium (2013, Prague). In 2018, he was awarded by the European Community (EU) the Marie-Sklodowska-Curie Individual Fellowship. He holds two patents and 3 patent applications in the area of acoustic resonators and RF systems. He is a technical reviewer for several journals, such as Applied Physics Letter, IEEE Transactions on Electron Devices, IEEE Transactions on Ultrasound, Ferroelectric and Frequency Control, IEEE Journal of MicroElectroMechanical devices, IEEE Electron Device Letter, Journal of Micromachine and Micro-Engineering, Journal of Applied Physics, IEEE Sensors Letter and Review of Scientific Instruments.